\documentclass[useAMS,usenatbib]{mn2e}
\usepackage[pdfpagemode={UseOutlines},bookmarks,bookmarksopen,colorlinks,citecolor={blue},urlcolor={red}]{hyperref}
\usepackage[flushleft]{threeparttable}
\usepackage{booktabs,caption}
\usepackage{graphicx}
\usepackage{amsmath}
\usepackage{subfigure}
\usepackage{amssymb}
\usepackage{tabularx}
\usepackage{natbib}
\usepackage{float}
\usepackage{etoolbox}

\usepackage{natbib}
\usepackage[usenames,dvipsnames]{xcolor}

\usepackage[normalem]{ulem}

\newcommand{\bea}{\begin{eqnarray}}
\newcommand{\eea}{\end{eqnarray}}

\makeatletter

\patchcmd{\NAT@citex}
  {\@citea\NAT@hyper@{%
     \NAT@nmfmt{\NAT@nm}%
     \hyper@natlinkbreak{\NAT@aysep\NAT@spacechar}{\@citeb\@extra@b@citeb}%
     \NAT@date}}
  {\@citea\NAT@nmfmt{\NAT@nm}%
   \NAT@aysep\NAT@spacechar\NAT@hyper@{\NAT@date}}{}{}

\patchcmd{\NAT@citex}
  {\@citea\NAT@hyper@{%
     \NAT@nmfmt{\NAT@nm}%
     \hyper@natlinkbreak{\NAT@spacechar\NAT@@open\if*#1*\else#1\NAT@spacechar\fi}%
       {\@citeb\@extra@b@citeb}%
     \NAT@date}}
  {\@citea\NAT@nmfmt{\NAT@nm}%
   \NAT@spacechar\NAT@@open\if*#1*\else#1\NAT@spacechar\fi\NAT@hyper@{\NAT@date}}
  {}{}

\makeatother

\newcommand{\gae}{\lower 2pt \hbox{$\, \buildrel {\scriptstyle >}\over {\scriptstyle \sim}\,$}} % % \gtrsim 

\newcommand{\lae}{\lower 2pt \hbox{$\, \buildrel {\scriptstyle <}\over {\scriptstyle \sim}\,$}} % % \lesssim 

\newcommand{\msun}{\ensuremath{\,\textrm{M}_{\odot}}}

\title[The dearth of NSCs in bright galaxies]{The dearth of nuclear star clusters in bright galaxies}

\author[Arca-Sedda, Capuzzo-Dolcetta and Spera]{M. ~Arca-Sedda$^{1}$\thanks{E-mail: m.arcasedda@gmail.com}, R. ~Capuzzo-Dolcetta$^{1}$ and M. ~Spera$^{2}$\\
$^{1}$Dept. of Physics, Sapienza, University of Rome, Piazzale Aldo Moro 5, I-00185, Rome (Italy)\\
$^{2}$INAF - Osservatorio Astronomico di Padova, Vicolo dell' Osservatorio 5, I-35122 Padova, Italy
}

\begin{document}
\date{Revised to 09-2015}

\pagerange{\pageref{firstpage}--\pageref{lastpage}} \pubyear{2015}

\maketitle

\label{firstpage}

\maketitle

\begin{abstract}
We investigate the interaction of a massive globular cluster (GC) with a super massive black hole (SMBH), located at the centre of its host galaxy, by means of direct $N$-body simulations. 
The results show that tidal distortions induced by the stellar background and the SMBH act on a time shorter than that of dynamical friction decay for a $10^6$ M$_\odot$ GC whenever the SMBH mass exceeds $\sim 10^8$ M$_\odot$. This implies an almost complete dissolution of the infalling GC before it reaches the inner region ($\lesssim 5$ pc) of the parent galaxy. 
The generalization of this result to a larger sample of infalling GCs shows that such destructive process may prevent the formation and growth of a bright galactic nucleus. Another interesting, serendipitous, result we obtained is that the close interaction between the SMBH and the GC produces a ``wave'' of stars that escape from the cluster and, in a fraction, even from the whole galaxy. 
\end{abstract}


\section{Introduction}
\nocite{*}

A galactic nucleus is a region where various astrophysical phenomena co-exist. 
Thanks to the high resolution images provided by the Hubble Space Telescope, it is clear,
nowadays, that the nuclei of the majority of elliptical and early type spiral galaxies (with mass 
$\gtrsim10^{10}M_\odot$) harbor massive and supermassive black holes (SMBHs) 
\citep{UrPa,FerrFord,Graham11,Shankar09,Kormendy13,Merri13,graham15}.
SMBHs have masses from $\sim 10^6\msun$, up to the extreme case of the galaxy NGC1277, 
which hosts a SMBH with mass $\sim 1.7\times 10^{10}$ M$_\odot$ \citep{VdeBo12}, although this mass estimation is very uncertain \citep{Ems13}.

In some cases, the SMBH is surrounded by a nuclear star cluster (NSC), a massive and dense stellar system composed of up to $10^8$ stars \citep{carollo97,boker02,cote06,wehner06,graham07,graham09}. In general, fainter galaxies host a NSC without evidence of the presence of a central massive black hole.
Hence, it seems that there is a continuos sequence from NSC-dominated galaxies to SMBH-dominated galaxies \citep{bekki10}. Due to this, SMBHs and NSCs are often indifferently referred to as compact massive objects (CMOs).

NSCs are observed in galaxies of any type in the Hubble sequence \citep{rich,carollo97,BKR02,cote06,Turetal12,denB} and 
their channels of formation and evolution are still matter of debate.
What we know, at present, is that NSCs are very massive ($10^6-10^7$ M$_\odot$) \citep{walcher05}, 
with a half-light radius of $2-5$ pc \citep{geha02,boker04}, and much more luminous ($\sim$4 mag) than ordinary 
galactic GCs.

Generally, NSCs contain both an old stellar population, age $ \gtrsim 1$ Gyr, and a younger one, with ages below $100$ Myr
\citep{rossa}.
Furthermore, NSCs are located at the photometric and kinematic centre of their host galaxy, i.e. at the bottom of its potential 
well \citep{BKR02,NEUM11}. 

Small sizes and large masses make NSCs the densest stellar systems observed in the Universe  
\citep{Neum12}.

Two are the most popular, and debated, formation scenarios: 

\begin{itemize}
\item the one commonly referred to as ``\textit{in-situ model}'', relies upon several injections of gas that accretes onto the central SMBH, leading to the formation and growth of a NSC
\citep{King03,Mil04,King05,McLgh,bekki07,nayakshin,hopkins10,Antonini15}; 

\item the scenario which is usually known as ``\textit{dry-merger scenario}''. In 
this case, the main engine is the action of dynamical friction that causes the sink 
of massive GCs toward the galactic centre \citep{Trem75,Trem76,Dolc93,antonini13,ASCD14a}.
Decaying clusters merge in the central galactic region leading to the formation, and subsequent growth, of a NSC.
This formation channel has been studied by several authors, through $N$-body simulations \citep{DoMioB, DoMioA, AMB, ASCD15He}.
\end{itemize}

Currently, it is very difficult to discriminate between the two presented formation scenarios. However, the existence 
of scaling relations between the CMOs and their host galaxies may provide additional clues on the modes of growth and evolution of NSCs and SMBHs.

In particular \cite{frrs} provided a power law relation between the NSC mass, $M_{\rm NSC}$, and the host velocity 
dispersion, $\sigma$, that is $M_{\rm NSC}\propto \sigma^4$. The in-situ model reproduces fairly well this correlation \citep{antonini13}.

On the other hand, a more recent analysis based on larger datasets, provided a shallower power law relation, with exponent between $2$ and $3$
\citep{LGH, ERWGD, graham12, scot}; this shallower relation is in good agreement with the predictions of the dry-merger model \citep{antonini13, ASCD14b}.

An interesting feature of NSCs is that they are very rare in galaxies with masses $\gtrsim 10^{11}$M$_\odot$. These galaxies have luminosity profiles that do not show observational evidences of a central excess, which is widely considered a clear signature of the presence of a NSC \citep{cote06,Turetal12,denB}.

Various hypotheses have been raised to explain the observed absence of a nucleated region in massive galaxies. 
For instance, giant elliptical galaxies are thought to be the merging product of smaller galaxies. In this framework, if two colliding galaxies contain both a SMBH and a NSC, the central region of the merger product should borrow a SMBH binary, that may heat the surrounding nucleus, inducing its evaporation \citep{Merri,bekki10,Antonini15b}. 

Another intriguing possibility is that the central SMBH prevents the NSC formation by tidally destroying its ``building blocks'', 
i.e. the decaying star clusters \citep{Dolc93,antonini13,ASCD14b}.
As a consequence of the tidal stripping exerted on the decaying clusters, it is also possible that the newly born NSC
is too small and, thus, very hard to detect through the analysis of the galactic luminosity profile.
 
On the other hand, as suggested recently by \cite{ASCD15He}, the tidal action of a SMBH seems insufficient to quench the formation of NSCs in galaxies containing SMBHs with masses similar (or less than) that of the Milky Way.

Combining the scaling relations that link NSCs and SMBHs with their host galaxies,   
it is found that the SMBH mass threshold over which galaxies do not show any evidence for a nucleated region roughly corresponds to the
condition $M_{\rm SMBH}\gtrsim M_{\rm NSC}$. As a matter of fact, no NSCs are observed in galaxies harboring very massive BHs at their centres; 
this could be the indication of a physical connection between the presence of NSCs and the SMBHs.

In this paper, we investigate the tidal action exerted  by a SMBH and by the galactic region in which it moves, on an infalling GC, which is supposed to be one of the building blocks of a NSC.

The main aim of this research is understanding whether the tidal disruption process can inhibit the formation of a NSC or, at least, limit its mass (and luminosity) below an observationally detectable value.

To reach such a goal, we performed a set of direct $N$-body simulations of the inner region of a galaxy containing a central SMBH and an orbiting GC.

The paper is organized as follows: in Section \ref{sec2} we present and discuss how we modelled the host galaxy and the GC, and the initial conditions selected for the simulations; our results are presented in Section \ref{sec3} and discussed in Section \ref{sec4}. Finally, in 
Section \ref{sec5}, we draw the conclusions of this work.

\section{Modelling the host galaxy and the orbiting cluster}
\label{sec2}

Using a series of high-precision, direct $N$-body simulations, we modelled the orbital decay of a single GC traversing the inner region of a galaxy containing a SMBH, whose mass is tightly correlated with the galaxy mass. We varied the SMBH and galaxy mass, in order to highlight how their tidal action influences the infall process of the GC.

In particular, we aim to shed light on which is the dominant process in shaping the inner region of galaxies: the dynamical friction, which drags GCs toward the galactic centre, making the formation of a NSC easier, or the tidal heating mechanism, which favours the disruption of the infalling GCs.

\subsection{The host galaxy}

A one-to-one $N$-body modelling of an entire galaxy requires an exceedingly large number of particles ($N\gtrsim 10^{11}$ for a typical galaxy).
Nowadays, the state-of-the-art of direct $N$-body simulations limits such number to $N\sim 10^{6}$.

In order to keep the number of particles reasonably low without altering the correct behaviour of the system, avoiding spurious relaxation effects and ensuring a sufficient phase-space resolution, we restricted our model to a limited spatial region of the galaxy. To generate a dynamically stable model of the spatial region of interest, we adopted a truncated density profile, that, as suggested by \cite{McMD}, has the form

\begin{equation}
\rho_{\rm tr}(r)=\frac{\rho(r)}{\cosh(r/r_{\rm tr})},
\label{truncdens}
\end{equation}

where $\cosh$ is the hyperbolic cosine function, $r_{\rm tr}$ is the truncation distance, 
that we assume $r_{\rm tr}=70$ pc, and $\rho(r)$ 
belongs to the class of the so-called \citep{Deh93} profiles:

\begin{equation}
\rho(r)=\frac{(3-\gamma)M_g}{4\pi r_s^3}
\frac{1}{(r/r_s)^{\gamma}(1+r/r_s)^{4-\gamma}},
\label{dens}
\end{equation}
where $M_g$ is the galaxy mass, $r_s$ is its length scale and $\gamma$ gives the inner slope of the density profile.

The density models considered in this paper have $\gamma$ in the range $[0.2,0.3]$. In particular, we selected $\gamma$ in order to obtain a good compromise between the reliability of the galaxy model and the resolution of its $N$-body sampling. Following this strategy, we used more than $1.5\times 10^4$ particles to model the infalling GC, allowing a reliable description of its internal dynamics.

For the purposes of this work, we sampled 4 galaxy models with mass in the range 
$10^{10} $M$_\odot<M_g<3.2\times 10^{11} $M$_\odot$.

The mass of the SMBH hosted at the centre of the galaxy was set according to the scaling relation provided by \cite{scot}:

\begin{equation}
{\rm Log}_{10} \left(\frac{M_{\rm BH}}{{\rm M}_\odot}\right) =\alpha {\rm Log}_{10} \left(\frac{M_g}{10^{11.3}{\rm M}_\odot}\right) + \beta,
\label{BHscal}
\end{equation}

where $\alpha = 1.37\pm 0.23$ and $\beta = 8.47\pm 0.07$. 

Table \ref{simula} summarises the main parameters that characterise our galactic representations.

\begin{table*}
\caption{}
\centering{Parameters of the simulations}
\begin{center}
\begin{tabular}{ccccccc}
\hline
\hline
$M_{g}$ &$M_{\rm BH}$ &$r_s$ &$M_{\rm tr}$ &$\sigma$ &$N_{\rm gal}$ &$m_*$\\
$(10^{11}$M$_\odot)$ & $(10^8$M$_\odot)$ & $({\rm pc})$ & $(10^7$M$_\odot)$ & $({\rm km/s})$ & & $($M$_\odot)$\\
\hline
$0.1  $&$ 5\times 10^-2  $&$995 $&$3.4$&$30$ &$1,018,742$&$34$\\
$0.32  $&$ 2\times 10^-1  $&$1512$&$4.1$&$40$ &$1,024,025$&$41$\\
$1.0  $&$ 1              $&$1917$&$5.9$&$70$&$1,031,338$&$57$\\
$3.2  $&$ 5              $&$2876$&$6.8$&$140$ &$1,033,332$& $66$\\
\hline
\end{tabular}
\end{center}
\begin{tablenotes}
\item Col. 1: galaxy mass, Col. 2: BH mass. Col. 3: galaxy scale radius. Col. 4:
galaxy mass within $r_{tr}$. Col. 5: velocity dispersion of the galaxy within $50$ pc from the BH. Col. 6: number of particles used in the galaxy.
Col. 7: individual particle mass. 
\end{tablenotes}
\label{simula}
\end{table*}

\subsection{The globular cluster model}

The phase space distribution of the stars in the GC is generated according to the King distribution function \citep{King}.

The mass of the GC model used in our simulation is $M=10^6$ M$_\odot$ since only such massive GCs have had time to sink toward the central galactic region, where we start our simulations.  

The tidal radius, $R_t$, gives an estimate of the  cluster size, and is evaluated with the relation \citep{bt}:

\begin{equation}
R_t^3=\frac{GM}{\omega_p^2+\displaystyle{\left(\frac{\mathrm{d}^2\Phi}{dr^2}\right)_{r_p}}},
\label{tid}
\end{equation}

where $r_p$ is the perigalactic distance of the GC, $\omega_p$ is the 
angular velocity of the circular orbit of radius $r_p$, and $\Phi(r)$ is the galactic gravitational potential, assumed spherically symmetric. 

Equation \ref{tid} is obtained assuming the cluster as a point of mass $M$. In our simulations, this hypothesis is acceptable because the typical core radius of the cluster ($R_c$) is much smaller than $r_p$, being $R_c/r_p \sim 0.03$.

The galactic potential, $\Phi(r)$, is given by the sum of the central BH potential and the smooth galactic background. Hence, the tidal force per unit mass acting on the GC at its pericentric distance is:

\begin{equation}
\left(\frac{\mathrm{d}^2\Phi}{dr^2}\right)_{r_p} = \frac{2GM_{\rm BH}}{r_p^3}-\frac{GM_g}{r_p^\gamma(r_p+r_s)^{4-\gamma}}\left[(1-\gamma)r_s-2r_p\right].
\label{dudr}
\end{equation}
The angular velocity $\omega_p$ in Equation \ref{tid} is:

\begin{equation}
\omega_p^2 =  \frac{GM_{\rm BH}}{r_p^3} + \frac{GM_g}{r_p^3}\left(\frac{r_p}{r_p+r_s}\right)^{3-\gamma}.
\label{omega}
\end{equation}

Equations \ref{tid}, \ref{dudr} and \ref{omega} make clear that the heavier the galaxy and the SMBH, the smaller the value of $R_t$ at fixed values of the other parameters.
In our simulations, the smaller value of $R_t$ is achieved in the galaxy model characterised by M$_{\rm BH}=5\times 10^8$ M$_\odot$ and $r_p\simeq 7$ pc, where $R_t \simeq 4$ pc.

The tidal radius of a GC modelled with a King profile is tightly connected to the adimensional potential well, $W_0$, and the core radius, $R_c$. To set these parameters, we ran a series of test simulations at varying $W_0$, $R_c$. We found that for $W_0\lesssim 6$ and $R_c>0.24$ pc, the GC is almost completely disrupted in less than a dynamical time, quite independently of the value of $R_c$. Hence, we choose for our GC model these two limiting values ($W_0 = 6$ and $R_c=0.24$ pc), avoiding this way spurious tidal effects on the GC evolution. 

In order to highlight the effects of the galactic nucleus and the SMBH on the orbital evolution of the GC, we decided to adopt the same GC model in all the simulations performed. 

In our simulations, we followed the dynamical evolution of the GC, placed on circular, radial and eccentric orbits, starting at an initial distance $r_0=50$ pc from the central SMBH. 
We stopped our simulations when either the GC distance to the SMBH falls below $20$ pc or the GC has lost $80\%$ of its initial mass.
With the choice of parameters given above, our GC models have 2-body relaxation times ($t_{\rm rel}$) exceeding tens of Myr in all the cases studied, thus implying evaporation times, $t_{\rm ev} \simeq 140 t_{\rm rel}$ \citep{bt}, much longer than the simulated time, which does not pass 130 Myr. 
We set the orbital parameters such that $r_0$ represents the initial apocentre in all the cases studied. 
This choice leads to orbits of different total energy, $E_0$, going from radial to circular orbits. However, the differences we found are smaller than $1\%$, as shown through the ratio $E_0/\Phi_(r_{\rm tr})$, being $\Phi(r_{\rm tr})$ the potential energy evaluated at the truncation radius (see Table \ref{simula2}).

We choose such small orbits around the galactic centre  because we want to study the effect 
of single or multiple close encounters between a massive GC and the central SMBH. From the computational point of view, the small orbital apocentres allow a good compromise between the 
numbers of particles used to model the GC and the galaxy, keeping a good level of resolution. It is 
worth noting that the BH influence radius encloses the GC orbit in the models with $M_{\rm g}>10^{10} 
$M$_\odot$.

Table \ref{simula2} reports a synthetic outline of the orbital properties of our GC.

\begin{table*}
\caption{}
\centering{GC orbital parameters}
\begin{center}
\begin{tabular}{cccccccccc}
\hline
\hline
$M_{g}$&$r_p$&$e$&$E_0/\Phi(r_{\rm tr})$&$v_0/v_c(r_0)$&$N_{clu}$\\
$(10^{11}$M$_\odot)$&$({\rm pc})$\\
\hline
$0.1$ &  $ 50$ &$ 0$&$ 1.038$&$1$&$29,832$\\
$0.1$ &  $ 7.4$   &$ 0.75$&$ 1.035$&$0.5$\\
$0.1$ &  $ 0$  &$ 1$&$ 1.025$&$0$\\
$0.32$ & $ 50$&$ 0$&$ 1.036$&$1$&$24,550$\\
$0.32$ & $ 7.4$&$ 0.74$&$ 1.031$&$0.5$\\
$0.32$ & $ 0$&$ 1$&$ 1.017$&$0$\\
$1.0$ & $ 50$&$ 0$&$ 1.053$&$1$&$17,237$\\
$1.0$ & $ 7.4$&$ 0.74$&$ 1.044$&$0.5$\\
$1.0$ & $ 0$&$ 1$&$ 1.018$&$0$\\
$3.2$ & $ 50$&$ 0$&$ 1.106$&$1$&$15,243$\\
$3.2$ & $ 7.8$&$ 0.73$&$ 1.087$&$0.5$\\
$3.2$ & $ 0$&$ 1$&$ 1.106$&$0$\\
\hline
\end{tabular}
\end{center}
\begin{tablenotes}
\item Col. 1: galaxy mass. Col. 2: pericentre of the orbit. Col. 3: eccentricity. Col. 4: ratio between the initial GC orbital energy and the potential energy at the truncation radius. Col. 5: ratio between the initial velocity of the cluster and the circular velocity at its initial position. Col. 6: number of particles used to model the GC.
\end{tablenotes}
\label{simula2}
\end{table*}

\section{Results}
\label{sec3}

All our simulations have been run using \texttt{HiGPUs} \citep{Spera}, a highly parallel, direct summation, $N$-body code that fully exploits the computational power of Graphic Processing Units (GPUs).

The \texttt{HiGPUs} code implements a Hermite 6-th order integrator with block time steps and individual softening, $\varepsilon$, to smooth the gravitational interactions among the stars within the GC.
In particular, we used $\varepsilon=10^{-3}$ pc. This 
choice leads to a relative error on the mechanical energy in the range $10^{-6}-5\times 10^{-5}$ over the whole time extension of the simulations.

After several experiments, we found that a total number of particles $N\gtrsim 10^6$ was a good compromise between phase space resolution of the whole system (galaxy+SMBH+GC), and computing time, allowing us to carry out a wide set of simulations in a reasonable time.
To study how the GC structure evolves as it moves around the galactic centre, we developed an analysis tool that provides, along the GC trajectory, estimates of the size and mass of the cluster. The determination of a centre for the GC is not a trivial task to accomplish, because the GC, during its motion, may be severely warped by the tidal forces. Hence, the simple use of the centre of mass (COM) can be unsatisfactory for an estimates of the GC mass, spatial size and determination of its position in the galaxy. This convinced us to develop a recursive, grid-based algorithm that allows  a reliable determination of the GC centre of density (COD), which permits a better evaluation of its structural and orbital parameters.

\begin{figure}
\centering
\includegraphics[width=8cm]{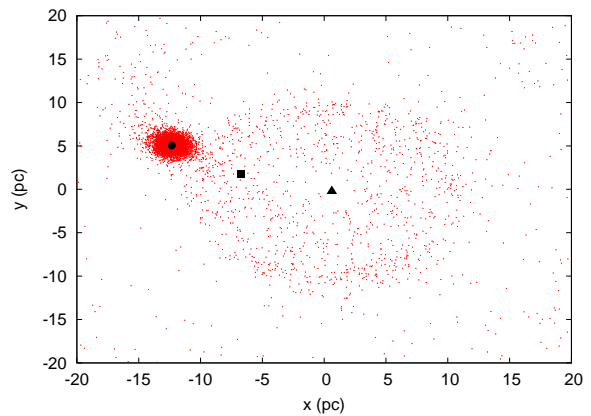}
\caption{One snapshot of the simulation characterised by $M_g = 3.2\times 10^{10} $M$_\odot$. The red dots represent the stars in the GC, while the filled black triangle 
identifies the position of the BH. The filled black circle indicates the COD of the cluster, while the open black circle represents its COM.}
\label{COD}
\end{figure}

Figure \ref{COD} shows the difference between the COM and the COD in the simulation with $M_g = 3.2\times 10^{10} $M$_\odot$; it is clear the importance in determing the actual centre of the cluster in a proper way.

\subsection{Circular Orbits}

The investigation of the effects of the interaction between the GC and the central SMBH has been done in the case of four different values of the galaxy mass, namely $M_g = 10^{10}, 3.2\times 10^{10}, 10^{11}, 3.2\times 10^{11} 
\rm{M}_\odot$, containing a SMBH, whose mass, obtained with Equation \ref{BHscal}, is reported in Table \ref{simula}.

In this Section, we discuss the results of the simulation of the evolution of a GC, placed on an initial circular orbit, at a distance $r_0 =50$ pc from 
the galactic centre. The galactic centre in these cases coincides with the position of the SMBH, since we didn't found significant displacement of the SMBH during the evolution. Hence, in the following we assume that the galactic centre coincides with the position of the SMBH.

\begin{figure}
\centering
\includegraphics[width=8cm]{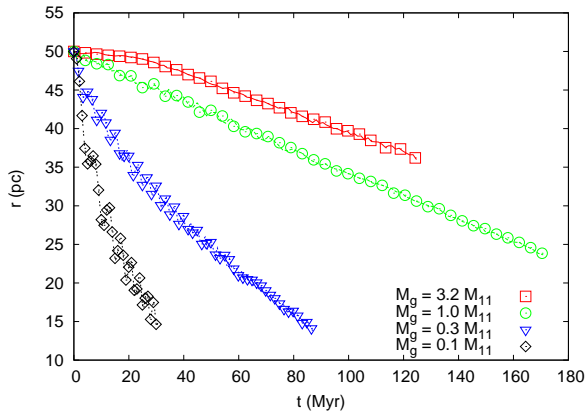}
\caption{Time evolution of the GC distance to the BH in the four cases studied, as labelled in the plot.}
\label{F3}
\end{figure}

Figure \ref{F3} illustrates the GC distance from the SMBH, as a function of the time. It is evident that galaxies hosting lighter SMBHs allow the GC to reach an inner galactic region, whose size is comparable to the typical dimension of a NSC. On the other hand, heavier galaxies, harboring SMBHs more massive than $\sim 10^8$M$_\odot$, efficiently shatter the GC, avoiding its decay to the innermost galactic region.

\begin{figure}
\centering
\includegraphics[width=8cm]{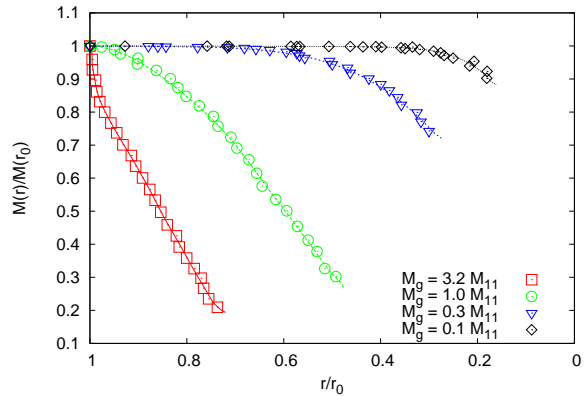}
\caption{The ratio of the GC mass to its initial value as function of its distance to the central massive black hole, in the case of an initially circular GC orbit. Each curve refers to a galactic model as labelled.}
\label{F4}
\end{figure}

Figure \ref{F4} shows the mass of the GC, normalized to its initial value, as a function of the istantaneous distance of the GC from the SMBH. This plot, combined with Figure \ref{F3}, allows to distinguish which mechanism dominates the GC evolution. Indeed, when tidal heating is the most effective phenomenon we should observe a rapid decrease of $M(r)/M(r_0)$ while $r/r_0$ decreases smoothly. On the other hand, if dynamical friction acts more efficiently than tidal heating, a nearly constant value of $M(r)/M(r_0)$ accompanies the $r/r_0$ decrease. The figure shows that tidal heating progressively deplete the GC of its stars for hosting galaxies more massive than $10^{11}$ M$_\odot$. 
On the other hand, the GC remains bound in lighter galaxies, and reach the inner region of the galaxy keeping more than $70\%$ of its initial mass.  
Table \ref{masslos} reports the radial distance of the GC from the galactic centre and the percentage of mass that remains bound to the GC at the end of our runs, making clear what stated above. The simulations ended at different times, since radial orbits decay faster than more roundish ones. Hence, the GC final mass listed in the table refers to different times.

Our results suggest that in a galaxy of mass above $\sim 10^{11}$ M$_\odot$ the combined bulge-SMBH tidal forces strip away from the GC most of its mass before it GC loses enough orbital energy to reach the inner galactic region. Conversely, below such {\it critical} value dynamical friction dominates over tidal heating, dragging the GC very close to the galactic centre.

\begin{table}
\caption{}
\centering{GC final mass}
\begin{center}
\begin{tabular}{ccc}
\hline
\hline
$M_g $ & $r_{\rm dec} $ & $M_{\rm dep}/M$ \\
$(10^{11}$M$_\odot)$ & $({\rm pc})$ &   ~(\%)\\
\hline
$0.1$           &$ 12$ &$ 89$ \\
$0.3$ &$ 12$ &$ 70$ \\
$1.0$           &$ 25$ &$ 27$ \\
$3.2$ &$ 35$ &$ 15$ \\
\hline
\end{tabular}
\end{center}
\begin{tablenotes}
\item Column 1: mass of the host galaxy. Column 2: distance of the GC from the SMBH  after the completion of the orbital decay process. 
Column 3: mass percentage which is still bound to the GC at the end of the simulation.
\end{tablenotes}
\label{masslos}
\end{table}

\subsection{Eccentric Orbits}

Given the definition of orbital eccentricity as $e=(r_a-r_p)/(r_a+r_p)$, where $r_a$ is the apocentric distance, after the circular ($e=0$), we investigated also eccentric 
($e >0$) and radial ($e = 1$) orbits to understand how the orbital type 
influences the transport of matter toward the galactic centre influencing the process of formation of a stellar nucleus therein.

In the case of radial orbits, the cluster is placed at $r_0=50$ pc from the SMBH with zero initial velocity, 
while eccentric orbits are characterized by the GC starting at $r_0$ with an initial velocity $v = 
v_c(r_0)/2$, where $v_c(r_0)$ is the circular velocity at the galactocentric distance $r_0$. The direction of the initial velocity is orthogonal to the radius vector, leading to an eccentricity $e\simeq 0.7$.

In these eccentric cases, the analysis of the GC structure is complicated in models with $M_g>10^{11}\rm{M}_\odot$, due to the difficulty to distinguish a clear centre and boundary for the GC after its first passage at pericentre.

\begin{figure*}
\centering
\includegraphics[width=16cm]{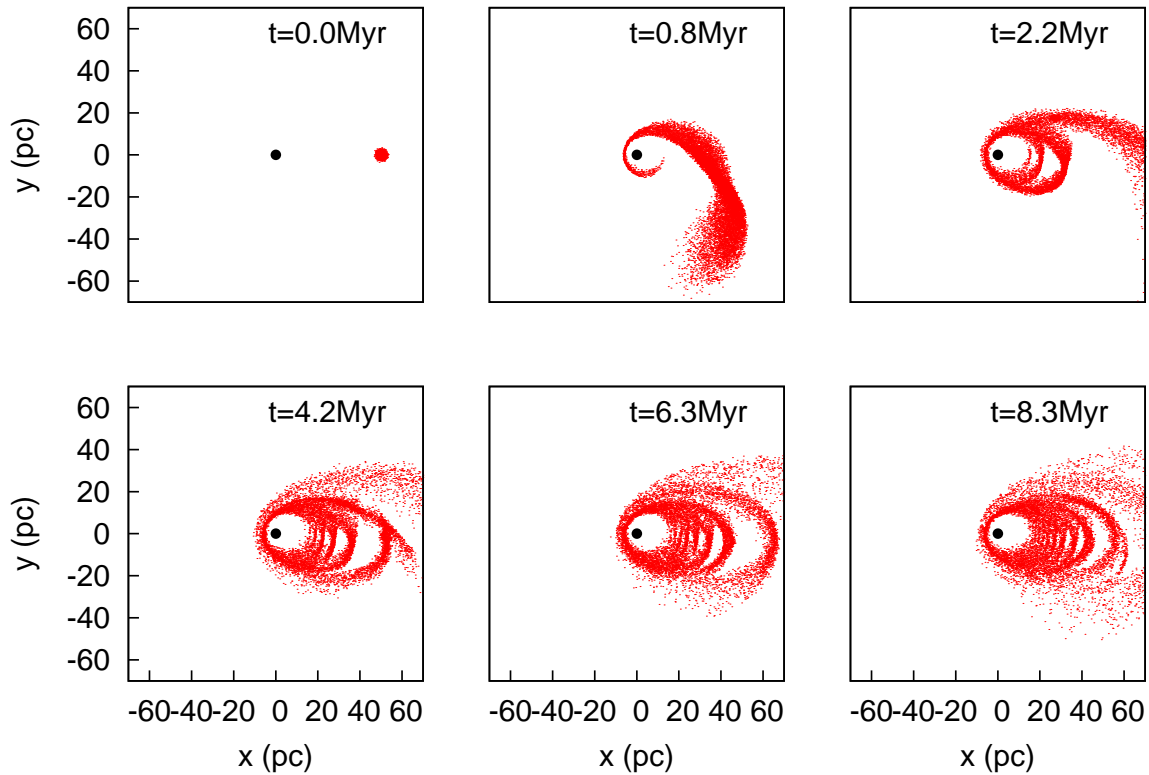}
\caption{Snapshots at different times (labelled) of the cluster on an eccentric orbit in the case $M_g=3.2\times 10^{11}\rm{M}_\odot$. The central black dot represents the BH with  mass $M_{\rm BH}=5\times10^8\rm{M}_\odot$. The x-y plane is the GC orbital plane.
}
\label{F5}
\end{figure*}

\begin{figure*}
\centering
\includegraphics[width=16cm]{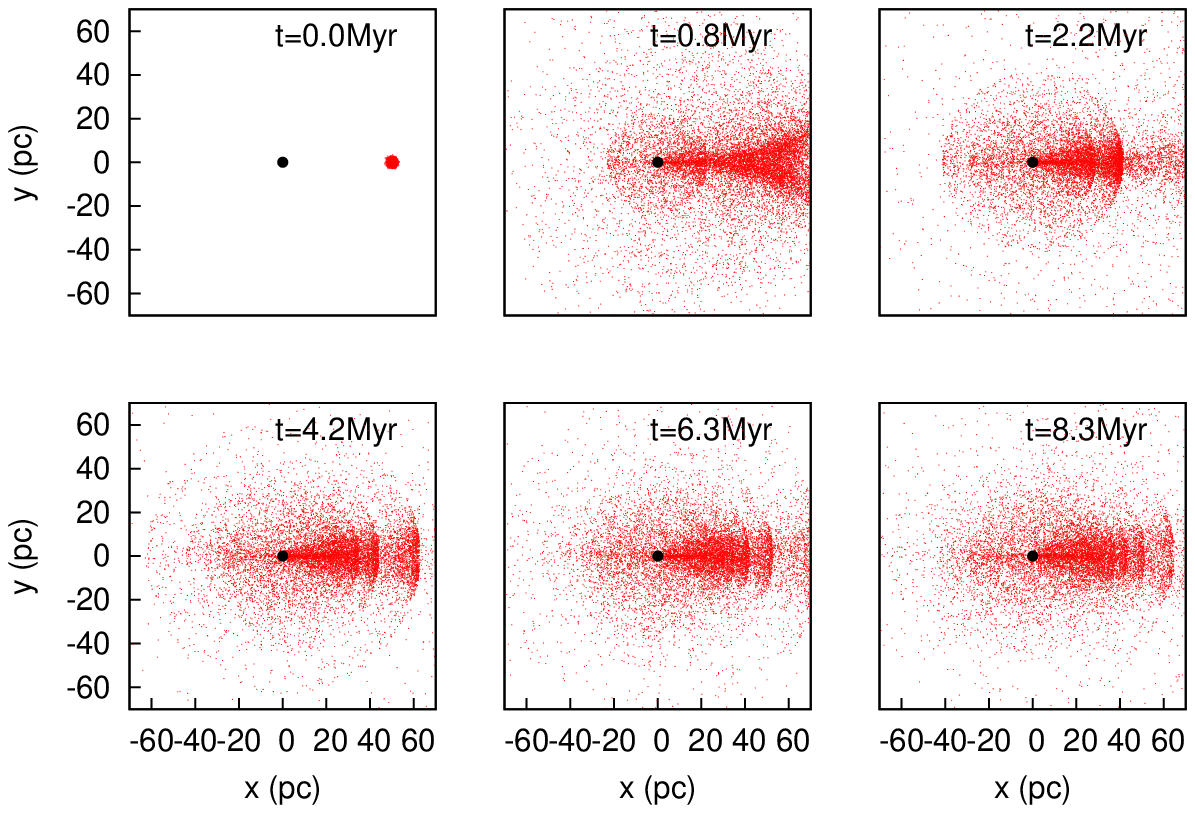}
\caption{Same as in Figure \ref{F5}, but for a cluster on a radial orbit.
}
\label{F6}
\end{figure*}

Figures \ref{F5} and \ref{F6} show some snapshots of the GC orbital evolution for eccentric and radial orbits in the case $M_g = 3.2 \times 10^{11}$ M$_\odot$. In the case $e=0.7$, it is worth noting that at every passage at pericentre some stars are stripped away from the GC and tend to move on precessing ellipses whose semi-major axes increase with the number of passages at pericentre. On the other hand, when the GC experiences an ``head-on'' collision with the SMBH, stars are scattered backward at different velocities, depending on the impact parameter that characterises each star-SMBH interaction. This peculiar interaction leads to the formation of a wake behind the former GC, with a structure similar to the propagation of a wave on a water plate. 

Therefore, although it is quite difficult to identify the ``core'' of the GC, the interaction between an SMBH and an infalling GC produces debris that hides much informations about the initial orbital parameters of the GC.

However, due to the fact that it is almost impossible to identify a centre for the cluster in these cases, we limited our analysis to the estimate of the amount of GC mass which remains confined to an inner region of the galaxy, around the SMBH.

When considering the simulations with $M_{\rm g}\geq 10^{11}$M$_\odot$, we found that eccentric and radial orbits 
allow a more efficient transport of mass, toward the galactic centre, than circular orbits do. On these orbits, the GC can carry about $20\%$ of its initial mass to the galactic centre, while, on circular orbits, such percentage is limited to few $\%$. 
On the other hand, for lower galaxy masses, $M_{\rm g}<10^{11}$M$_\odot$, we found that the major contribution to the deposited mass comes from GC on circular orbits. 
We will deepen the discussion about these results in Section \ref{sec4}.

\section{Discussion}
\label{sec4}

In this paper we have presented several $N$-body simulations 
to study the interaction between a single GC and a SMBH in the galactic centre. The results  shed light on the actual mechanisms of mass accumulation in 
the central region of a massive galaxy ($M_g \ge 10^{11}$ M$_\odot$).

\begin{figure}
\centering 
\includegraphics[width=8cm]{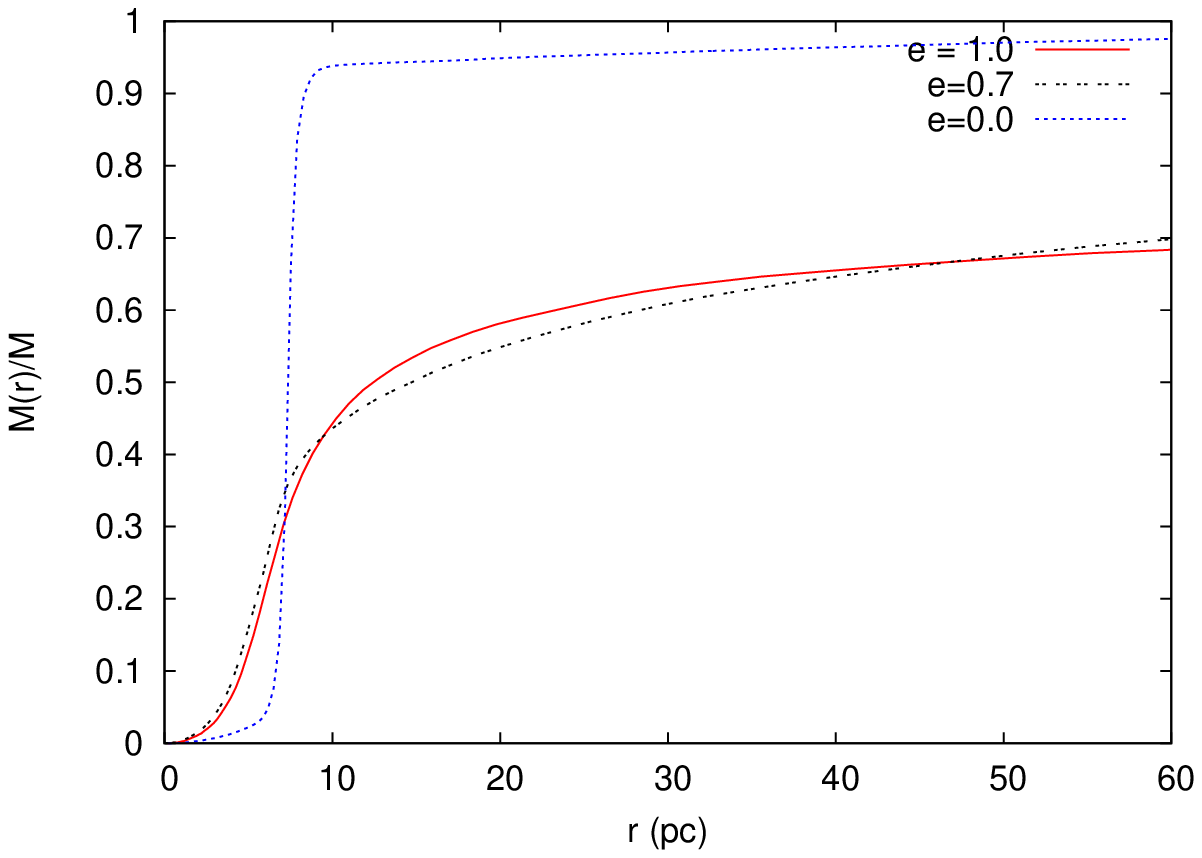}\\
\includegraphics[width=8cm]{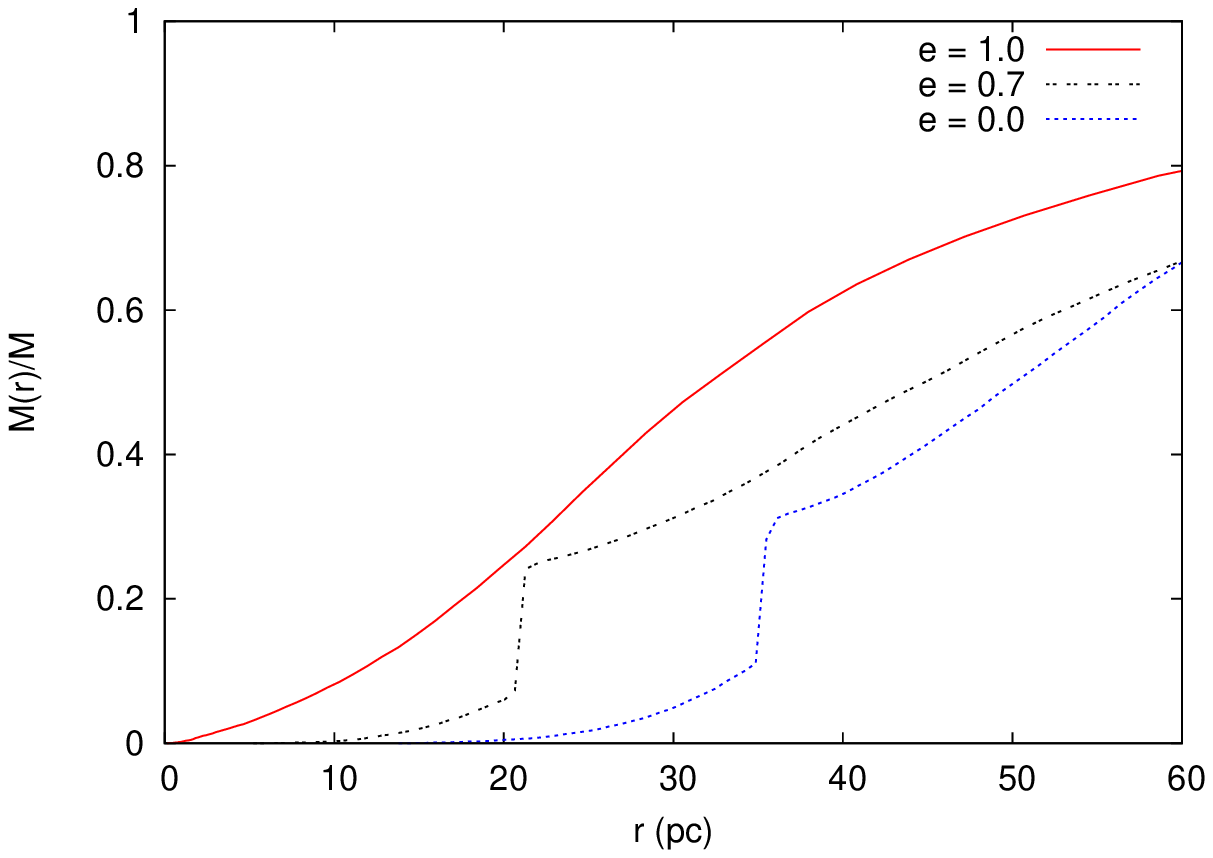}\\
\caption{Cumulative mass profile of the GC, normalised to its initial mass, at the end of the simulations. The galaxy models considered here are $M_g=10^{10}$ M$_\odot$ (top panel), and $M_g=3.2\times 10^{11}$ M$_\odot$ (bottom panel). In each panel, the various lines refer to GC orbits of the intial eccentricity labeled on top.}
\label{F14a}
\end{figure}

Figure \ref{F14a} shows the GC cumulative mass profile as a function of the distance from the SMBH after the  orbital decay process, for all the orbits considered and in the case of $M_g = 10^{10} $M$_\odot$ and $M_g = 3.2\times 10^{10} $M$_\odot$. Considering  our lightest galaxy model, we found that circular and radial orbits are both efficient in the transport of mass toward the galactic centre. In such cases, the mass left to the galactic centre exceeds $80\%$ of the initial mass of the GC.
On the other hand, in heavier galaxy models, only GCs moving on initial nearly-radial orbits deposit a non negligible amount of their mass ($\sim 20\%$ of the initial mass of the GC) in the inner $20$ pc of the galaxy.  
Hence, the fraction of mass deposited within $20$ pc from the SMBH varies from few percent, for the heaviest galaxy model, to more than $90\%$ in the lightest.

We have shown that the mass deposition around the galactic centres is not efficient in galaxies with masses above $M_g=10^{11} $M$_\odot$. However, we cannot exclude that such (unefficient) mechanism may, eventually, drive the formation of a detectable, central overdensity. In order to address this point, we tried to reconstruct the projected, combined density profile of the host galaxy and its orbitally decayed GCs.

To do this, we consider a population of $N_{\rm GC}$ clusters whose orbits are evenly distributed among circular, eccentric ($e\simeq 0.7$) and radial.

Under the (extreme) assumption of ``linearity'' of the decay process, we can evaluate the global density profile as the mere sum of the density profile of the galaxy, $\rho_g(r)$, and of the decayed clusters:
\begin{equation}
\rho(r)=\rho_g(r)+\sum_{i=1}^3\alpha_i\rho_{e_i}(r),
\end{equation}
where $\alpha_i$ represent the fraction of GCs with initial eccentricity $e_i$. In the following calculations, we used $\alpha_i = 1/3$, and $e_1=0$, $e_2=0.7$, $e_3=1$.

\begin{figure}
\centering 
\includegraphics[width=8cm]{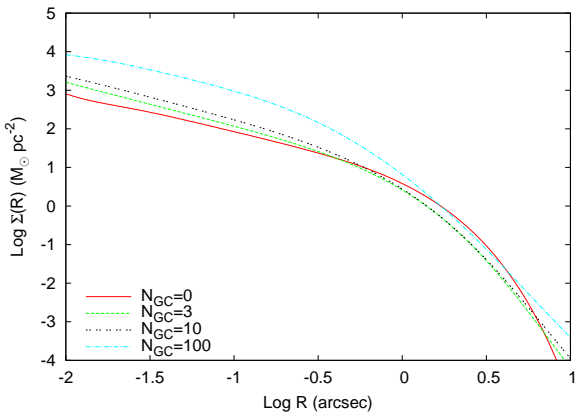}
\caption{Projected density profile of the galaxy (red line) compared to the density profile in the cases of full decay of different numbers of GCs ($N_{\rm GC}$) as labelled.
}
\label{denpro}
\end{figure}

It should be noted that the detectability of a nucleus through the analysis of the projected density profiles depends on the resolution of the instrument used to look at the target galaxy and the distance of the target itself. Due to this, in Figure \ref{denpro} we assumed the host galaxy at the distance of the Virgo cluster (taken as $16.5$ Mpc \citep{Mei07}), where a quite large number of NSCs have been detected over time \cite{cote04,cote06,frrs}.

Generally, the presence of a NSC in a galactic centre is argued by the presence of a clear edge in the projected inner luminosity profile of the host galaxy. Hence, a parameter that can be used to discriminate wheter or not a NSC resides in the centre of a galaxy is the surface density contrast ${\delta \Sigma}/\Sigma$ between the total (galaxy+GC) and the galactic background. 
We found that in our simulations the surface density contrast is related to the number of orbitally segregated clusters, $\delta \Sigma/\Sigma \propto N^k$, with $k\simeq 0.87$. 

Galaxies with a clearly visible nucleated region have typical values $\delta \Sigma/\Sigma \sim 10$ 
\citep{cote06,denB}. On the base of our simulations we deduce that such values can be achieved only when more than $100$ clusters with masses $\sim 10^6$ M$_\odot$ reach the galactic centre within a Hubble time.

Consequently, the missing evidence of NSCs in a galaxy of mass $M_g=3.2\times 10^{11}$M$_\odot$ would mean that in such galaxy there have not been enough GCs to merge and form it.
This result agrees with previous findings by \cite{ASCD14b} indicating that in a galaxy with that mass the number of clusters which can decay and contribute to the formation of a nucleus is less than $\sim 100$. 

Hence, the disruptive mechanism proposed here represents a satisfactory alternative to other theories, such as the disruption of pre-existing NSCs after a major merger event, or the tidal heating caused by a massive BH binary at the centre of the host galaxy. It is worth highlighting that our proposed mechanism requires only two ingredients, well supported by theoretical and observational arguments, that are: i) the presence of a SMBH within the galactic centre, and ii) the presence of a populations of star clusters around it. Moreover, it involves time-scales which are significantly shorter than those expected for galaxy merging and BHB formation.

\section{High velocity stars from GC-MBH encounters}

The simulation of the interaction of the GC with the central massive galactic black hole showed an interesting side effect: after the fly-by around perigalacticon, a certain amount of stars are extracted from the GC and some of them even from the galaxy, on a privileged direction along the GC trajectory. 

In the galaxy  hosting the most massive BH, $M_{\rm BH}=5\times 10^8$ M$_\odot$, after the passage at 
perigalacticon, some stars in the GC gain velocities up to $\sim 850$ kms$^{-1}$. A fraction of them leave the galaxy reaching distances up to $\sim 15$ kpc in $60$ Myr and (nearly constant) residual velocities up to $\sim 250$ kms$^{-1}$.

We performed a careful investigation to check whether it was a real effect or due to errors related to the numerical integration. This was done comparing the simulation results with those coming from extremely accurate integrations 
(although with a reduced number of particles) performed using a serial, fully regularized version of our N-body code, where regularization is applied to the whole set of pair interactions. Its use allowed us to keep the variation of the total mechanical energy of the system below $10^{-13}$.
This code employs the Mikkola's algorithmic regularization \citep{mikkola99,preto99}, coupled with the chain algorithm, in order to handle efficiently very large particle mass ratios, like those we have in this work (massive black hole respect to the star mass). 
Results of this code (applied to a subsample of objects) confirmed that the effect was real and not a numerical artifact.
Actually, we simulated the orbital evolution of the GC in the case $M_{\rm BH}=5\times 10^8$ M$_\odot$. Since our fully regularized, serial, $N$-body is hugely time consuming, we simulated only the first passage at the pericentre of our GC, sampled with $\sim 50$ particles. We found that the fraction of stars that leave the cluster and their velocities agree very well with the values obtained in the ``extended'' case.

It is worth noting that the escape velocity from the whole system, evaluated in our most massive model at perigalacticon, is, in our most massive model, $v_e(r_p)\leqslant 771$ kms$^{-1}$, while the escape velocity from the cluster is $\sim 90$ kms$^{-1}$. 
Hence, a fraction of the total number of stars are possible candidate as escapers, i.e. as stars energetically unbound from the galaxy and the GC.

\begin{figure}
\centering
\includegraphics[width=8cm]{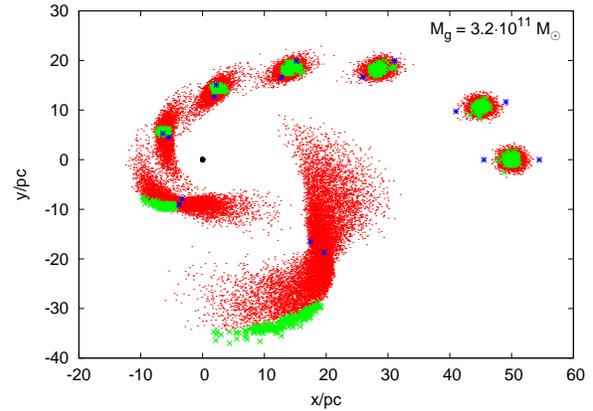}
\caption{Various snapshots of the GC moving in a counter clockwise motion on an eccentric orbit. Escaping stars are green crosses, while red dots identify the stars that remain bound to the cluster. The black filled circle labels the SMBH, while the blue asterisks represent the lagrangian points L1 and L2. After the passage at pericentre, stars are thrown through L2.}
\label{F7}
\end{figure}

In Figure \ref{F7} we present some snapshots of the GC moving in the $M_g=3.2\times 10^{11}$M$_\odot$ galaxy 
on the $e\simeq 0.7$ orbit, through its first perigalacticon passage, marking in color the escaping stars. 
After the passage at the pericentre, stars flow away through the lagrangian point L2, along the direction tangential to the GC trajectory. Therefore, the ejection occurs on a preferential direction, leading the escapers to move in a sort of ``collimated beam''.

\begin{figure}
\centering 
\includegraphics[width=8cm]{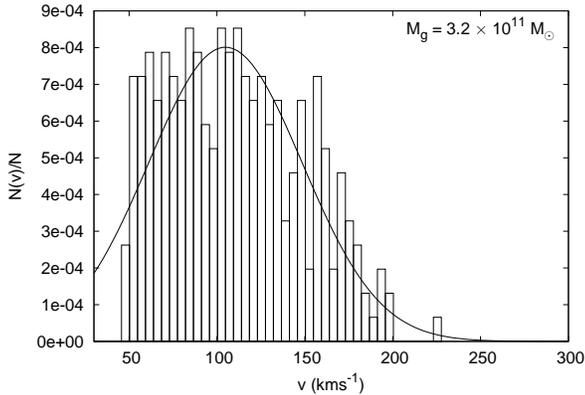}
\caption{Velocity distribution of escapers $63$ Myr after the GC passage at perigalacticon for a cluster containing $10^6$ stars.  All the stars in the plot have distance from the galaxy centre larger then $1$ kpc. The black line indicates a gaussian fit with mean value $103\pm3$ km s$^{-1}$ and dispersion $46.5\pm3.8$ km s$^{-1}$.}
\label{F122}
\end{figure}

We identified as escapers from the galaxy those stars that both i) get a positive mechanical energy in the inertial system of reference, and ii) reach galactocentric distances exceeding $10^2$ times the truncation radius of our galaxy model, thus making extremely implausible a recapture.
Using this procedure, we identified 362 escaping 
stars in the galaxy hosting the $5\times 10^8$ M$_\odot$ massive BH, that is $2.4\%$ of the total number of stars of the GC. 
Scaling this number to the total number of stars of a realistic GC, say $10^6$ stars, we obtain a population of $\sim 24,000$ stars,  
formerly belonging to the GC, which wander in the outer region of the galaxy, with velocities in the range $50-250$ kms$^{-1}$ and positive total energy.
Figure \ref{F122} shows the velocity distribution of the whole population of escaping stars. 

 We found that the escapers have residual velocity up to $250$ km s$^{-1}$, a value comparable to the velocities of high-velocity stars detected in the Milky-Way \citep{blaauw,poveda,hobbs,brown,silva,brown12,brown14}, 
but definitively smaller than that of the hyper-velocity stars, whose velocities 
can exceed $800-1000$ km s$^{-1}$ \citep{hills,yutremaine,gualandris05,sesana06,rossi}.

Recently, \cite{brown14} have shown that most of the high-velocity stars observed in our galaxy are located in a preferential region, near the galactic North Pole. Of course, this anistropic distribution may hidden some clues related to their origin.
 
Intriguingly, we notice here that the interaction between an infalling GC and a SMBH seems to represent a significant channel of formation of high-velocity stars, producing a collimated beam of stars that, in a certain fraction, can even escape from the host galaxy. Hence, this channel deserve further investigations \citep{cd15}.

Likely, the correct interpretation of the velocity gain is by mean of a 3-body interaction among the GC, the SMBH and the generic star of the cluster. 
The basic concept, recently developed in \cite{cd15}, is that some stars of the cluster during its fly-by around the SMBH gain enough kinetic energy, subtracted to the GC-SMBH pair, to be expelled from the GC and even from the galaxy.  

\begin{figure}
\centering
\includegraphics[width=8cm]{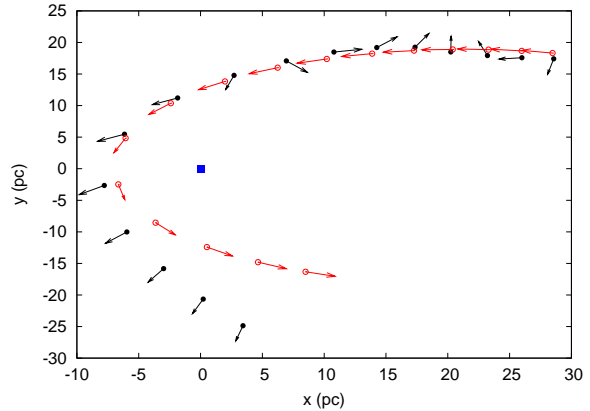}
\caption{Production of a high-velocity star (black dot) in our simulation. The black vector represents the velocity  of the escaping star while the red vector identifies the velocity of the GC centre of density. The  blue square is the central SMBH.}
\label{3body}
\end{figure}

Figure \ref{3body} shows the velocity vector of one of the escaping stars in simulation $M_g=3.2\times 10^{11}$ M$_\odot$, as it moves around its GC, which, in its turn, is moving around the SMBH. After the passage at pericentre, it is evident that the star, which is behind the GC with respect to the GC-SMBH direction, is accelerated and thrown away along the direction tangential to the GC orbit.

\begin{figure}
\centering 
\includegraphics[width=8cm]{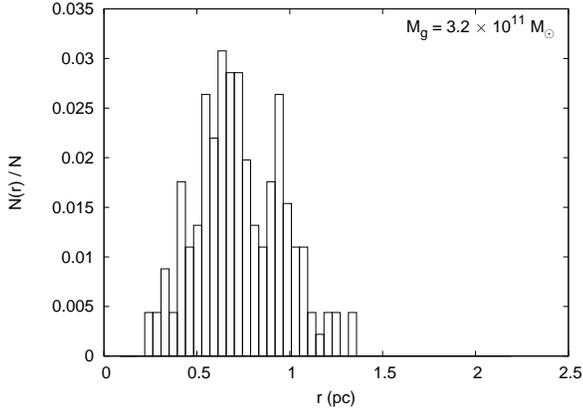}
\caption{Spatial distribution, with respect to the GC COD, of the escapers that gain the largest kicks during the passage at pericentre. We considered here only stars with velocities greater than the escapers mean velocity, $\simeq 103$ km s$^{-1}$. 
The distribution, fitted with a gaussian distribution, peaks at $r = 0.71\pm 0.24$ pc.}
\label{rhysto}
\end{figure}

This 3-body mechanism is similar to the interaction between the SMBH and a binary star proposed by \cite{hills}, but in this case the ``primary'' member of the binary is the whole GC, or at least the fraction of the cluster mass enclosed within the trajectory of the ``secondary'' (the future escaping star). 
Using the parallelism with the treatment of \cite{hills}, later revisited by \cite{yutremaine} and \cite{rossi}, we can give a rough estimate of the ejection velocities of stars gaining kinetic energy at distance $r_*$ from the GC center, that, in the case of a GC Plummer profile gives

\begin{equation}
v_{\rm ej}({\rm km ~s}^{-1})= 108 \left(\frac{M_{\rm BH}}{10^6 ~{\rm M}_\odot}\right)^{\frac{1}{6}} \left(\frac{M}{10^6 ~{\rm M}_\odot}\right)^{\frac{1}{3}} \left(\frac{1 ~{\rm pc}}{r_P}\right)^{\frac{1}{2}}
\left(\frac{x_*}{1+x_*^2}\right)^{\frac{1}{2}}
\label{ejPlu}
\end{equation}

which maximizes for $x_*=r_*/r_P=1$, where $r_P$ is the profile scale length. 
During the passage at pericentre, the GC half-mass radius for our model is $r_h = 0.8 \pm 0.2$. For a Plummer sphere, $r_h \approx 1.3 r_P$; therefore, for our GC model $r_P = 0.6 \pm 0.1$ pc, value very close to the peak of the velocity distribution in one of our $N$-body simulation (see Fig. \ref{rhysto}) which can be considered quite representative of simulations presented in this paper.

\section{Conclusions}
\label{sec5}

In this paper we gave an interpretation of the observed dearth of NSCs in massive host galaxies. 
The problem was studied by performing several high-precision, direct, $N$-body simulations of the orbital decay of a massive GC	 in a galaxy harboring a central 	SMBH. To run our simulations we employed the \texttt{HiGPUs} code \citep{Spera}. These simulations allowed us to quantify the combined role of dynamical friction and of tidal forces on the cluster along its motion. 

Our results can be summarized as follows:

\begin{itemize}
\item in all the cases studied (different masses for the host galaxy and for the central black hole, different initial orbits for the ``test'' GC, as summarized in Table \ref{simula}), we found that the main contribution 
to the ``tidal heating'' of the cluster was given by the central SMBH, accompanied by a less relevant and 
slower erosion caused by the stellar galactic background;

\item in galaxies below $10^{11}$M$_\odot$, the cluster  transports to the galactic centre more than $80\%$ of its initial mass, regardless of the shape of the initial orbit. On the other side, in heavier galaxies we showed that the  mass deposited to the central galactic region by the cluster on the nearly radial orbit is limited to $20\%$ of its initial  mass, while the cluster moving on the initially circular orbit is almost completely disrupted after its first  crossing of the galactic centre, thus giving a negligible contribution to the formation of a bright nucleus therein;

\item by means of scaling arguments we showed that in galaxies more massive than $10^{11}$M$_\odot$ the formation of a clearly detectable stellar projected overdensity (a NSC) should occur only when the total mass of the decayed clusters (of the size and with the characteristics studied in this paper) is $\sim 10^8$M$_\odot$, an order of magnitude above the value expected from observational and theoretical arguments.

\item in the case of GC eccentric and radial orbits we found that, as the cluster passes through the perigalacticon in the case $M_g=3.2\times 10^{11}$M$_\odot$, a small quantity of the cluster stars is ``thrown'' away from the galactic centre reaching distances above $5$ kpc with residual velocities up to $250$ kms$^{-1}$. A fraction of these stars reach velocities such to become unbound from the galaxy. Rescaling these results to a real globular cluster, with $\sim 10^6$ stars, this corresponds to more than $10^4$ escapers. The anisotropy in the escapers distribution and the velocities that they gained suggest that this sort of 3-body mechanism (GC+SMBH+star orbiting the GC) is a valid mode of formation of high-velocity stars. This deduction finds a nice confirmation in that the velocity distribution of the escapings stars in our $N$-body simulations peaks for stars receiving the kick when they transit at a distance from the GC center equal to the Plummer length scale of the analytical model on which we based our 3-body considerations, following the original \cite{hills} approach.

\end{itemize}

\section*{Aknowledgements}

The authors warmly thank S. Mikkola for providing his code implementing the algorithmic regularization, and for his useful comments and suggestions about the star-ejection mechanism discussed in this paper.

MAS acknowledges the MIUR, which funded part of this research through the grant PRIN PRIN 2010 LY5N2T 005, and the financial support from the University of Rome Sapienza through the grant D.D. 52/2015 in the framework of the research programme "The MEGaN project: Modelling the Evolution of GAlactic Nuclei". 

MS acknowledges financial support from the Italian Ministry of Education, University and Research (MIUR) through grant FIRB 2012 RBFR12PM1F.

\footnotesize{
\bibliographystyle{mn2e}
\bibliography{bblgrphy2}
}

\end{document}